\documentclass[printer]{aa}
\usepackage{graphicx}
\usepackage{txfonts}
\usepackage{natbib}
\bibpunct{(}{)}{;}{a}{}{,}

\begin{document}
\def\rappk24{$f(24\mu m)/f(K)$}

\title{The fraction of quiescent massive galaxies in the early Universe}

   \author{A. Fontana \inst{1}
   \and
   P. Santini \inst{1,2}
   \and
   A. Grazian \inst{1}
   \and
   L. Pentericci \inst{1}
   \and
   F. Fiore \inst{1}
   \and
   M. Castellano  \inst{1}
   \and
   E. Giallongo \inst{1}
   \and
   N. Menci \inst{1}
   \and
   S. Salimbeni \inst{1,3}
   \and
   S. Cristiani  \inst{4}
   \and
   M. Nonino \inst{4}
   \and
   E. Vanzella  \inst{4}
   }

   \offprints{A. Fontana, \email{fontana@oa-roma.inaf.it}}

\institute{INAF - Osservatorio Astronomico di Roma, Via Frascati 33,
00040 Monteporzio (RM), Italy \and Dipartimento di Fisica,
Universit\`{a} di Roma ``La Sapienza'', P.le A. Moro 2, 00185 Roma,
Italy \and Department of Astronomy, University of Massachusetts, 710 North Pleasant Street, Amherst, MA 01003 \and INAF - Osservatorio Astronomico di Trieste, Via G.B.
Tiepolo 11, 34131 Trieste, Italy }

   \date{Received .... ; accepted ....}
   \titlerunning{The fraction of quiescent massive galaxies in the early Universe}

   \abstract {} { 
     We attempt to compile a complete, mass--selected sample of
     galaxies with low specific star--formation rates, and compare their
     properties with theoretical model predictions.}
   { 
     We use the \rappk24 flux ratio and the SED fitting to the
     $0.35-8.0 \mu$m spectral distribution, to select quiescent
     galaxies from $z\simeq 0.4$ to $z\simeq 4$ in the GOODS--MUSIC
     sample.  Our observational selection can be translated into
     thresholds in specific star--formation rate $\dot{M}/M_*$, which
     can be compared with theoretical predictions. }
   { 
     In the framework of the well-known global decline in
     quiescent galaxy fraction with redshift, we find that a non-negligible
     fraction $\simeq 15-20\%$ of massive galaxies with  low
     specific star--formation rate exists up to $z\simeq 4$, including
     a tail of ``red and dead'' galaxies with
     $\dot{M}/M_*<10^{-11}$yr$^{-1}$. Theoretical models vary
     to a large extent in their predictions for the fraction of galaxies
     with low specific star--formation rates, but are unable to
     provide a global match to our data. }


\maketitle
%

\section{Introduction}

Understanding the formation and evolution of early--type
galaxies is a major goal of present-day cosmology, as well as a
fundamental benchmark for ``ab-initio'' theoretical models of galaxy
evolution.

According to several independent lines of evidence, the population of
massive galaxies has undergone major evolution during the epoch
corresponding to the redshift range $1.5-3$, where the galaxy stellar
mass function evolved significantly at high masses
\citep{berta07,fontana06} (F06 in the following), massive galaxies
settled onto the Hubble sequence (e.g., \citet{abraham07} and
\citet{franceschini06}), and the red sequence appears in high $z$
clusters \citep[e.g.,][]{zirm08}.

The nature of the physical processes responsible for this rapid rise
remains unclear.  A large number of massive ($\simeq
10^{11}M_\odot$) actively star--forming galaxies is clearly in place
at $z\simeq 2$ \citep{daddi04,papovich07}. Within this population, the more massive galaxies tend to be the more actively star--forming
\citep{daddi07}, at variance with trends measured in the local Universe.

At the same time, galaxies with low levels of
star--formation rates at $z\simeq 1.5-2$ have been detected by imaging
surveys based on color criteria \citep[e.g.,][]{daddi04} or SED fitting
\citep{grazian07},  and
by spectroscopic observations of red galaxy samples
\citep{cimatti04,saracco05,kriek06}. These results have motivated the
inclusion of efficient methods for providing a rapid assembly of
massive galaxies at high $z$ (such as starburst during interactions)
as well as quenching of the SFR, most notably via AGN feedback
\citep{menci06,bower06,hopkins07}.

Unfortunately, a detailed validation of the prediction of theoretical
models has been hampered so far by the lack of a statistically well
defined sample of early--type galaxies at high redshift, and by the
difficulty in defining a common criterium to identify early--type
galaxies in the data. It is difficult to isolate passively
evolving galaxies from the wider population of intrinsically red
galaxies at high redshift, which include also a (probably larger)
fraction of dust-enshrouded star--forming galaxies. The two
classes are indeed indistinguishable when selected by means of a
single color criterium, such as the ``ERO'' classification ($R-K>4$)
\citep{daddi00,mccarthy04} or the ``DRG'' one ($J-K>2$)
\citep{franx03,vandokkum03}.  The corresponding SEDs, however, are not
degenerate and can be distinguished with more complex criteria, even
when spectroscopy is not feasible.  Some of these criteria adopt more
colors, such as the $I-J / J-K$ method proposed by \cite{pozzetti00} and
spectroscopically validated by \cite{cimatti03}, or the $BzK$ method
proposed by \cite{daddi04}. Other methods rely on the spectral fitting
of the overall SED, either by making use of the resulting rest--frame
colors, as in the case of the $U-V / V-I$ criteria proposed by
\cite{wuyts07}, or directly using the output of the SED fitting
process \citep{arnouts07,salimbeni08,grazian07}.

We used the method of using SED fitting output in our  
analysis of the GOODS-S data set, to disentangle the different
contributors to the mass density at high redshift \citep{grazian07}
and to describe the evolution in the luminosity function of red
galaxies up to $z\simeq 3$ \citep{salimbeni08}.

In this work, we use additional information available from
observations in the mid-IR $24\mu$m band, to allow a more careful
selection of high redshift passively evolving galaxies, with the
specific aim of comparing their number density with theoretical
expectations.

We use a revised version of our GOODS-MUSIC sample \citep{grazian06},
a 15 band multicolor (U--to--24$\mu$m ) catalog extracted from an area
of 143.2 arcminutes squared within the GOODS--South public survey.
The key improvements compared to our previous work are a revised IRAC
photometry and the addition of  $24\mu$m flux data for all objects
in the catalog. Full details are given in a companion paper \citep[][
S09 in the following]{santini08}.  Another important difference is
that we include all objects detected in the $4.5\mu$m image, down to
$m_{45}<23.5$. In the following, we adopt a mass-selected sample,
obtained by applying a mass threshold at $M_*\geq 7\times
10^{10}M_\odot$ to our photometric catalog based on a combined selection
$K<23.5$ {\it or} $m_{45}<23.5$. This photometric sample is complete
at this mass limit to $z\simeq 4$,  also for dust--absorbed star--forming
galaxies (see F06).

Observed and rest--frame magnitudes are in the AB system,
and we adopt the $\Lambda$-CDM concordance  model
($H_0=70km/s/Mpc$, $\Omega_M=0.3$ and $\Omega_{\Lambda}=0.7$).


\section{Quiescent galaxies}

\begin{figure}
\includegraphics[width=10cm]{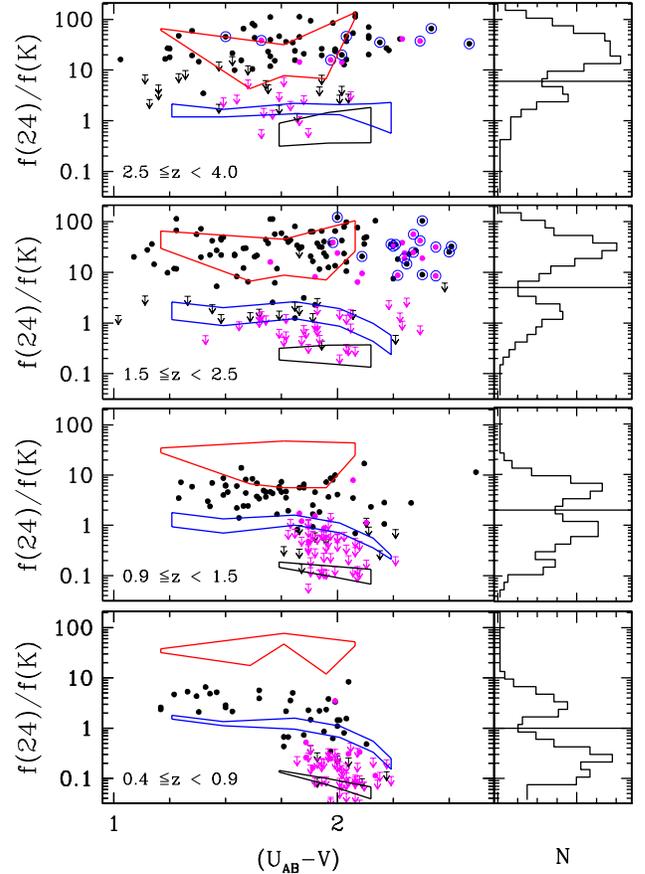}
\caption{ Ratio of the observed fluxes at $24\mu$m to that in the $K$
  band as a function of the rest-frame $U-V$ color, in four redshift
  bins, for the $M_*\geq 7\times10^{10}M_\odot$ sample. Upper limits
  correspond to objects fainter than $20\mu$Jy in the $24\mu$m
  image. Galaxies with $age/\tau>6$ (as discussed in Sect.
  3) are shown in magenta. Large blue circles are the ``obscured AGN''
  candidates of \citet{fiore08}. Closed areas represent the range of
  $f(24\mu m)/f(K)$ obtained by redshifting a set of local samples
  \citep{polletta07}: from bottom to top, early--type galaxies (black
  line), spirals (blue line), starbursts (red line).  }
\label{K24flux}
\end{figure}

\subsection{The role of mid-IR observations}
In principle, the observed mid--IR flux is a powerful tool for
distinguishing between the two classes of red galaxies at high
redshift. Dust--absorbed star--forming galaxies are expected to be
bright in the mid--IR, where most of the UV--light absorbed by dust is
re--emitted. In contrast, passively evolving galaxies are expected to
be far dimmer, since ordinary stars have low emission at IR wavelengths.
To quantify this criterium, we use the ratio \rappk24 between the
observed flux at $24\mu$m and the $K$--band flux.  We show the \rappk24
for all galaxies in our sample in Fig. \ref{K24flux}, in four redshift
bins. To show how this color can help us to differentiate between the
galaxy types, we computed the same \rappk24 to $z\simeq 4$ for a set
of local templates including early--type, spiral, and starburst
galaxies \citep{polletta07}. All templates were $k$--corrected at the
different redshifts. As shown in Fig. \ref{K24flux}, the loci populated
by the different galaxy types in the $U-V$ versus \rappk24 plane are
separated well, allowing us to distinguish between actively
star--forming galaxies and those with moderate--to--low star
formation.

In Fig. \ref{K24flux}, we show the observed \rappk24 for the complete
$M_*\geq 7\times10^{10}$ $M_\odot$ sample, both for individual galaxies
as well as its general distribution.  Upper limits were assumed to be
equal to the $1\sigma$ upper limits in the $24\mu$m photometry. It is
clearly shown in Fig. \ref{K24flux} that the \rappk24 ratio of
star--forming galaxies increases with redshift, and that a significant
population of starburst exists at $z>1$, with colors similar to those of
local, rarer LIRGs.  On the other hand, few bright galaxies at low and
intermediate redshifts, with low star--formation rates are
detected at low levels in the $24\mu$m image.

The most striking result is that \rappk24 distribution is
clearly bimodal, its minimum corresponding to a gap between
the loci of active and passive galaxies.  The observed minima in the
distribution of the \rappk24 ratio occurs at $ f(24\mu m)/f(K) =
1,2,5$ and 6 in the four redshift bins adopted in Fig. \ref{K24flux}. Two
warnings are required about this bimodality. First, it is enhanced by the upper limits at faint $24\mu$m levels, the true
distribution probably being wider even at lower \rappk24. More
importantly, the bimodality disappears if a deeper catalog is
used (for instance, the $z\leq 26$ selected version of the GOODS--MUSIC
catalog). The high mass-limit of our sample, indeed, preferentially
selects extremely red galaxies, as shown by their significantly red $U-V$ color, which
are those for which the bimodality is clearly evident. This is a key
feature of our approach, since the \rappk24 ratio provides a natural
way of distinguishing between active and quiescent galaxies within the
population of high--redshift, red galaxies.

\begin{figure}
\includegraphics[width=7.5cm]{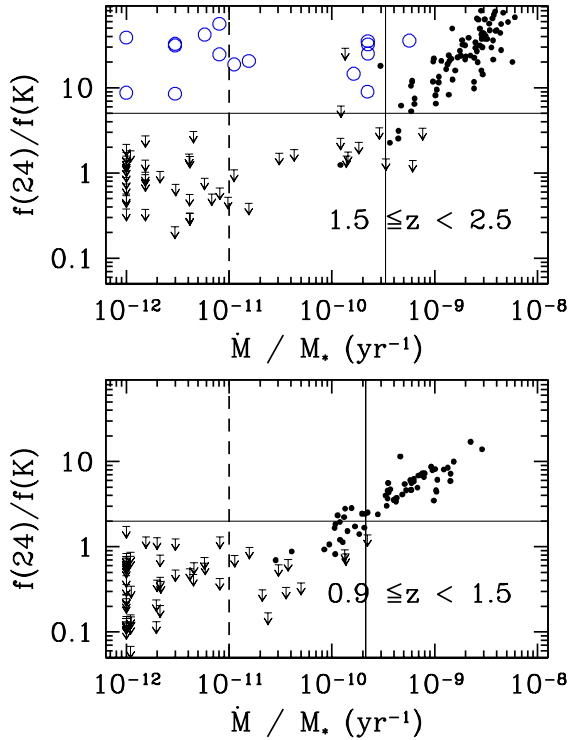}
\caption{ Ratio between the observed fluxes at $24\mu$m and in the $K$
  band as a function of the Specific Star--formation Rate
  ($\dot{M}/M_*$), for the sample with $M_*\geq
  7\times10^{10}M_\odot$. Galaxies with $\dot{M}/M_*<10^{-12}$ yr$^{-1}$
  have been arbitrarily set to $\dot{M}/M_*=10^{-12}$ yr$^{-1}$.  Upper
  limits refer to objects fainter than $20\mu$Jy in the $24\mu$m
  image.  Large blue circles are the ``obscured AGN'' candidates of
  \citet{fiore08}.  The solid vertical line corresponds to the inverse
  of the age of the Universe at the centre of the redshift bin. The
  dashed vertical line shows the threshold on $\dot{M}/M_*$ adopted to
  classify ``red and dead'' galaxies. The horizontal solid line refers
  to the minimum in the observed distribution of the \rappk24 ratio in
  the same redshift range. }
\label{DTime}
\end{figure}

\subsection{The selection on rest--frame quantities}
The criterium that we adopted to identify quiescent galaxies cannot be
directly applied to the output of current theoretical models for
galaxy formation and evolution, which typically do not predict the
mid--IR flux.  The comparison becomes more natural when we convert the
data into the quantities immediately provided by these models,
i.e., stellar mass or star--formation rates. We use the stellar
masses $M_*$ provided by the SED fitting, of accuracy described at
length in F06, and briefly in Sect. 3.  To derive star--formation
rates for galaxies detected at $24\mu$m, we apply the method of
\citet{papovich07} in converting the $24\mu$m + rest--frame UV luminosity
into a total star--formation rate, using the \citet{dalehelou02} models
with a (lowering) correction at high mid--IR fluxes. For galaxies
below $20 \mu$Jy at $24\mu$m, we use the SFR derived from the SED
fitting.  As we show in S09 \citep[see also][]{daddi07}, these two SFR
estimators agree relatively well, especially at low SFR levels.

In Fig. \ref{DTime}, we show the correlation between the \rappk24 ratio
and $\dot{M}/M_*$ in our sample.  For simplicity, we only plot the two
redshift ranges that are more populated in our sample: the same trend
holds at higher and lower redshifts. This correlation
is largely expected given the relationship between $\dot{M}$ and the
$24\mu$m flux on the one hand, and between $M_*$ and the $K$ magnitude on the
other.  The key point, however, is that the correlation is so tight
that it allows us to translate the (observational) criterium based on the
\rappk24 ratio (horizontal line in Fig. \ref{DTime}) into a
(model--oriented) cut to the estimated $\dot{M}/M_*$ (vertical line in
Fig. \ref{DTime}). As shown in Fig. \ref{DTime}, the two samples in
practice coincide.

Based on this evidence, we use the specific star--formation rate
$\dot{M}/M_*$ to select quiescent galaxies, which allows direct
comparison of the data with theoretical models. Given that $\dot{M}/M_*$ is
dimensionally the inverse of a timescale, a natural threshold of
$\dot{M}/M_*$ is the inverse of the age of the Universe at the
corresponding redshift $(t_U(z))^{-1}$.  We define galaxies with
$\dot{M}/M_*<(t_U(z))^{-1}$ as ``quiescent'' in the following. Such a
name is motivated by the fact that - by definition - these galaxies
have experienced a major episode of star--formation prior to the
observations \footnote{Indeed, if $M_* =
  \langle\dot{M}\rangle_{past}\times t_U(z)$, where
  $\langle\dot{M}\rangle_{past}$ is the star--formation rate averaged
  over the whole age of the Universe at the corresponding $z$, the
  requirement $\dot{M}/M_*<(t_U(z))^{-1}$ implies
  $\dot{M}\leq\langle\dot{M}\rangle_{past}$.}.

%

\section{``Red and dead'' galaxies}

\subsection{Basic definitions}

\begin{figure}
\includegraphics[width=9cm]{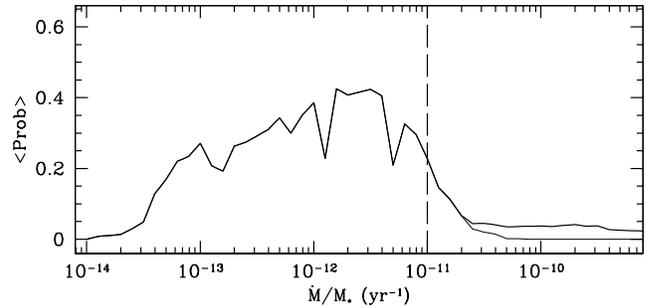}
\caption{ Probability distribution of the $\dot{M}/M_*$ ratio in the
  sample of ``red and dead'' galaxies at $1.5 \leq z \leq 2.5$,
  averaged over the whole sample. The thick curve shows the probability
  distribution of the SED fitting to the 14 bands SED, from the U band
  to the $8\mu$m band. The thin curve shows the probability
  distribution removing models that over predict the SFR with respect
  to the upper limit provided by the constraint at $24\mu$m. The
  dashed vertical line shows the limit $\dot{M}/M_*<10^{-11}$yr$^{-1}$
  that we adopt to define ``red and dead'' galaxies. }
\label{mdotm_error}
\end{figure}

Galaxies defined as ``quiescent'' by the criterium presented in the
previous Sect. may still have a measurable amount of ongoing SFR
(e.g., $\dot{M}\simeq 10 M_\odot/$yr for a $M_*\simeq10^{11}M_\odot$
galaxy).  It is therefore interesting to estimate the fraction of
galaxies with low or negligible levels of SFR. As we show in
the following, the physical properties of these objects, that we label ``red and dead'', constitute a sterner test to theoretical models.  

To select galaxies at high redshift on the basis of low star--formation rates, we need to complement our \rappk24 ratio data with output from the fitting analysis of the
optical--IR observed SED.  The lack of detection in the $24\mu$m image
provides only an upper limit to the ongoing SFR (see for
instance Fig. \ref{DTime}), and its exact level can be estimated from
the SED fitting only.  We applied the SED fitting technique
following the recipe described in several papers (see
\citet{fontana04} and F06 for details). The U-to-$8\mu$m photometry was compared with
a grid of models from the \citet{bc03} (BC03) spectral synthesis
code, characterized by exponentially declining star--formation
histories of timescale $\tau$ for a set of ages, metallicities, and
dust extinctions. 
For comparison with our previous work, and most of the
literature, we used the standard Salpeter IMF and the BC03 models.  We
verified however that adoption of the most recent version of
the code incorporating the treatment of the post--AGB stars
\citep{bruzual07} does not change our results significantly (see
Salimbeni et al. 2009 for a preliminary analysis). In particular, only
1 (out of 144) objects classified as ``red and dead'' would not be
classified as such with the new version.

To select ``red and dead'' galaxies, we use a 
threshold of $\dot{M}/M_*<10^{-11}$yr$^{-1}$ (vertical dashed line in
Fig. \ref{DTime}).  To some extent, this threshold is arbitrary, the
important point being that we adopt the same cut in
theoretical models, to ensure proper comparison. However, we note
that the same threshold was also used previously
\citep{brinchmann04} to separate active and quiescent galaxies. 
We also note that it corresponds to a threshold
$age/\tau>6$ between the fitted age and the star--formation
exponential timescale $\tau$. This follows from the fact that,
assuming exponentially declining star--formation histories, the
specific star--formation rate can be computed analytically to be
$\dot{M}/M_* = (\tau (e^{(age/\tau)}-1))^{-1}$, yielding
$\dot{M}/M_* \simeq 10^{-11}$yr$^{-1}$ for $age/\tau \simeq 6$,
for small values of $\tau$. This coincidence allows us to compare the
pure SED fitting approach with the additional analysis provided by the
$24\mu$m data.  A threshold to $age/\tau$ was applied before \citep{arnouts07,salimbeni08,grazian07} to broadly
identify passively evolving galaxies, although with a less
conservative value of $age/\tau>4$.  In Fig. \ref{K24flux}, we plot the location in the $U-V$ vs. \rappk24 plane of
all galaxies with $age/\tau>6$ (magenta points). 
 The classification scheme based purely on the optical--near IR
SED fitting agrees well with the \rappk24 criterium, providing an
important consistency check. The main exception are a number
of ``Compton thick AGN'' candidates at $1.5<z<2.5$, selected as described in
\citet{fiore08}, where the $24\mu$m emission was attributed to AGN
activity. These objects are discussed further in Sect. 3.4.

We select ``red and dead'' galaxies by requiring that
$\dot{M}/M_*<10^{-11}$yr$^{-1}$, not only that $age/\tau>6$. This is because when estimating SFR, we assign the SED--derived
values only to objects undetected in the mid--IR,
i.e., those with $24\mu$m flux below $20\mu$Jy. For this reason, the
few objects fitted with $age/\tau>6$ but with large \rappk24 are
not included in our sample of ``red and dead'' galaxies.  We now  discuss the accuracy of this method for the most
interesting subsample, i.e., ``red and dead'' galaxies at $z\geq
1.5$.



\subsection{``Red and dead'' galaxies at $1.5 \leq z \leq 2.5$}

A sizeable fraction of the sample with $\dot{M}/M_*<10^{-11}$yr$^{-1}$
have redshifts in the range $1.5 \leq z \leq 2.5$, where we detect 32
galaxies. 
The reliability of their $\dot{M}/M_*$ measure can be assessed using
the error analysis already widely adopted in similar
cases \citep[][F06]{papovich01}. Briefly, the full synthetic library
used to determine the best-fit model spectrum is compared with the observed
SED of each galaxy. For each spectral model (i.e., for each combination
of the free parameters $age$, $\tau$, $Z$ and $E(B-V)$) the probability
$P$ of the resulting $\chi^2$ is computed and retained, along with the
associated $\dot{M}$ and $M_*$.  The ensemble of $\dot{M}/M_*$ values
corresponding to probabilities above a given minimum threshold defines
the range of acceptable values for $\dot{M}/M_*$. Since most of our
``red and dead'' galaxies have only a photometric redshift, our analysis for these objects has been completed by allowing the redshift parameter to be free. In Fig. \ref{mdotm_error}, we show the resulting distribution of the
probability $P$, averaged over the entire sample of ``red and dead''
galaxies at $1.5 \leq z \leq 2.5$. The tail at large values of
$\dot{M}/M_*$ is due to 2 objects with a nearly equal probability of
being fitted accurately by a dusty starburst model solution. The correspondingly large values of $\dot{M}$ that would be required would easily overpredict the observed flux (which is an upper limit) at $24\mu$m.  Taking into account this additional
constraint, we display in Fig. \ref{mdotm_error} the distribution
of $P$ after removing the models that overpredict the SFR with respect
to the value provided by the upper limit in the emission at $24\mu$m.  It is clear from
Fig. \ref{mdotm_error} that the SED fitting for these galaxies is well
constrained at levels of $\dot{M}/M_*<10^{-11}$yr$^{-1}$, since the
probability of having $\dot{M}/M_*$ has only a small tail above
$10^{-11}$yr$^{-1}$.

This result is unsurprising. Since the galaxies in our sample are by
selection massive ($M_*\geq 7\times10^{10}M_\odot$), the threshold
$\dot{M}/M_*=10^{-11}$yr$^{-1}$ is equivalent to small but
 measurable amounts of SFR, of approximately $1
M_\odot$yr$^{-1}$. These levels correspond to detectable fluxes in the deep B-z optical images of GOODS (the magnitude computed from a
galaxy with $\simeq 1 M_\odot$yr$^{-1}$, for small values of $E(B-V)$,
is in the range 26-28 {\it mags}) and can then be measured by the full
SED fitting.

We can also check the small subset of our ``red and dead'' sample with spectroscopic observations in this redshift
range. Using both the K20 and the public GOODS data set, we found 6
objects that have a spectroscopic redshift. Four of these objects are
classified as early spectral type, with no detectable [OII] emission line. Two other
objects have both absorption features, typical of evolved systems, and a weak [OII] line. A weak [OII] line is not necessarily an
indicator of ongoing star--formation \citep{yan06}. Adopting the standard conversion of \cite{kennicutt98}, we derive star
formation rates of the order of 2-3 $ M_\odot$yr$^{-1}$, consistent
with the SED fitting estimates, confirming that these objects have
$\dot{M}/M_*\lesssim10^{-11}$yr$^{-1}$.  We conclude that our analysis
is sufficiently robust at $1.5<z<2.5$, and allows to build a mass--selected sample of ``red and dead'' galaxies.

\subsection{``Red and dead'' galaxies at $z > 2.5$}
\begin{figure}
\includegraphics[width=9cm]{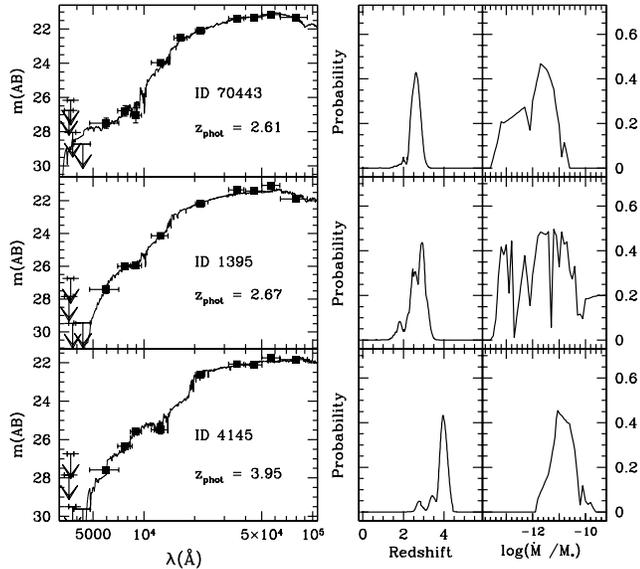}
\caption{ Examples of ``red and dead'' galaxies at $z>2.5$.For each
  object, from left to right: the observed flux in the GOODS band and
  the best-fit SED; the redshift probability function; the
  $\dot{M}/M_*$ probability distribution. From top to bottom, the three
  galaxies shown represents three different categories: a robust
  candidate at $2.5<z<3$; a less secure candidate in the same redshift
  range; a typical example of the candidates at $z>3$. }
\label{mdotm_z3}
\end{figure}

As we move to higher redshifts, $2.5 \leq z \leq 4$, our analysis
becomes more uncertain. First, the constraints on the $24\mu$m
emission are weaker or non--existent, since this band is shifted outside the range dominated by dust emission. 
Objects also become even fainter and redder, and in general lack any
spectroscopic confirmation. In this redshift range, we detect 12 ``red and dead'' candidates, 9 located at $2.5<z_{phot}<3$. We 
carefully examined the data for each object to assess the detection reliability. First, we verified that the photometry was not contaminated by nearby
companions, and that the overall SED was smooth and not biased obviously
 by photometric fakes. All galaxies selected in this redshift range
are very red with $z-B>4$, which is redder than typical
EROs at lower redshifts. Their optical--IR SED is dominated by a 
break between the $K$ and the IRAC bands, and a change of slope that is
a  signature of an evolved galaxy population (star--forming, dusty
objects exhibit a more featureless slope). The
redshift probability distribution is broad, but typically
implies that $z_{phot}>1.5-2$ with a relatively large spread $\Delta
z_{phot}\simeq 0.5$.

Quantitatively, the error analysis described above indicated that 5 of
the 9 objects at $2.5<z_{phot}<3$ are constrained to have
$\dot{M}/M_*<10^{-11}$yr$^{-1}$, even for such a broad redshift range. An example of these objects is shown in the upper panel of
Fig. \ref{mdotm_z3}. The four other  objects have a wider distribution of $\dot M / M$ (Fig. \ref{mdotm_z3}, middle panel), as well as 
redshift. 

In a similar way, the physical properties and the nature of the three objects detected at $z>3$ are weakly constrained. They have a clear minimum in the
$\chi^2$ distribution around $z_{phot}\simeq 3.5-4$, but have a tail to $z\simeq 2$. The distribution of acceptable
$\dot{M}/M_*$ values also extends significantly beyond $10^{-11}$yr$^{-1}$.

We conclude that at least 55\% of
our ``red and dead'' candidates at $2.5<z_{phot}<3$ are robust
candidates, while the remaining 45\% and the three candidates at $z>3$
are more uncertain. It would be natural to convert this finding into
an upper limit to the true number density of
``red and dead'' galaxies. However, we note that a comparable number of
galaxies are inferred to have $\dot{M}/M_*$ slightly
above $10^{-11}$yr$^{-1}$ and are not included in our
sample, but have a range of acceptable model fits extending well below $\dot{M}/M_*=10^{-11}$yr$^{-1}$. We conclude that
at $z>2.5$ the estimate of $\dot{M}/M_*$ becomes unreliable, due mainly to the limitations in the depth of existing
observations, and that the resulting statistical analysis of ``red and dead''
galaxies must be treated with caution.  As we show, the statistical error (including both Poisson and cosmic--variance components) is so large for these small samples that a more
rigorous treatment of these uncertainties is probably unnecessary.

\subsection{Hosts of Compton thick AGNs} 

A potential source of uncertainties in many statistical analysis of high
redshift galaxies is contamination by AGNs.  As we describe in
S09, we removed from our catalog both 
spectroscopically confirmed AGNs and X-ray sources with an
optically dominant point--like source. We also identified galaxies that likely harbor Compton thick AGNs using the
criterium defined by \citet{fiore08}, which is based on the detection
of a mid--IR excess in very red galaxies. This criterium is similar to
the one adopted by \citet{daddi07b} but also includes objects that are
much redder in the optical--IR range than the BzK--selected sample of
\citet{daddi07b}. A full discussion of the SED of these objects will
be presented elsewhere: for the moment, we note that 9 of these
``Compton thick AGN'' candidates at $1.5<z<2.5$ are classified as
``red and dead'' galaxies, as shown in Fig. \ref{K24flux}. Unfortunately,
the nature of these objects remains elusive.  Most of these
objects are among the reddest in our sample, as indicated by their
extreme rest-frame $U-V$. They are extremely faint or even
undetected in the deep $z$-band ACS images, and are included in our
sample only because of their detection in either $K$ or $4.5\mu$m
images. As a result, the probability distributions of their photometric
redshifts are often broad, although they are typically constrained to
be at $z_{phot}>2$.  Only three (of nine) galaxies have the
distribution of $\dot{M}/M_*$ constrained to be below $
10^{-11}$yr$^{-1}$. In the following, we do not rely on this
classification, and we therefore compute the fraction of
``red and dead'' galaxies at $1.5<z<2.5$ in three different ways (labelled {\it a), b), c)} in the caption of Fig. \ref{ActQuie}), to
explore all possible cases. First (option {\it a)}), we remove
all ``Compton thick AGN'' candidates from our sample, irrespective
of their SED classification. Alternatively (option {\it b)}), we assume that none of them is a ``red and dead'' galaxy, which
provides a lower limit to their fraction. Finally (option {\it c)}),
we include them in our statistical analysis, assigning their formal SED
classification. As we show in the following Sect., the fraction
computed with these different assumptions changes its value by $\simeq$20\%, still providing interesting constraints on theoretical
models.

\section{Results and discussion}

\begin{figure}
\includegraphics[width=9cm]{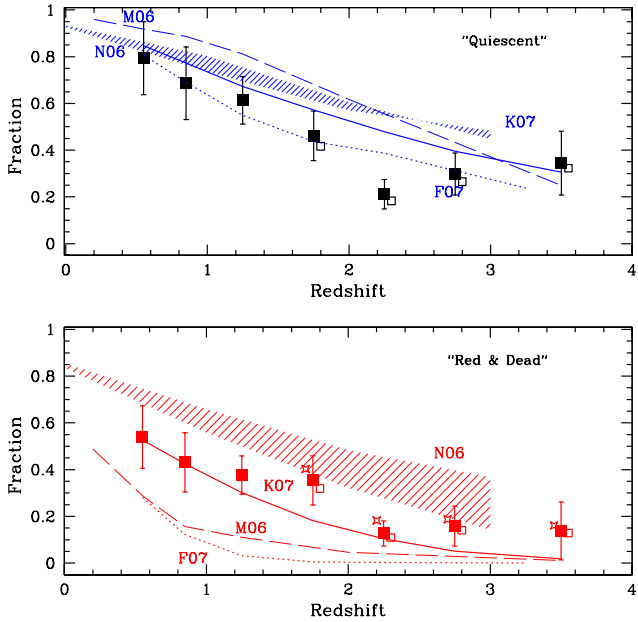}
\caption{ {\it Lower} Fraction of ``red and dead'' galaxies (defined by
  $\dot{M}/M_*<10^{-11}$yr$^{-1}$) as a function of redshift, in the
  $M_*\geq 7\times10^{10}M_\odot$ mass--selected sample. Points
  represent the observed values. Filled, open and starred points
  (slightly offset for clarity) refer to the three different
  strategies {\it a,b,c} discussed in the text to account for obscured
  AGNs.  Error bars include Poisson and cosmic variance errors.  Lines
  refer to the predictions of theoretical models, as described by the
  legend: \citet{menci06} (M06), \citet{fontanot07} (F07),
  \citet{kitzbichler07}(K07), \citet{nagamine06}, (N06); {\it Upper}
  Fraction of ``quiescent'' galaxies as a function of redshift,
  defined by $\dot{M}/M_*<(t_U(z))^{-1}$, in the same mass--selected
  sample. Points and lines as in the lower panel.  }
\label{ActQuie}
\end{figure}

We finally compute the fraction of ``quiescent'' and ``red and dead''
galaxies in different redshift ranges, following the definitions
provided in the last two Sections.  These fractions are shown in
Fig.\ref{ActQuie} as a function of redshift.  The different numbers of
``red and dead'' galaxies at $1.5<z<2.5$ reflect the different accounting methods of hosts of Compton thick AGNs.  Error bars were computed by summing (in quadrature) the Poisson and cosmic variance error. The latter was computed by measuring the relative variance
within 200 samples bootstrapped from the Millennium Simulation
\citep{kitzbichler07}, using an area as large as GOODS-S and applying
the same selection criteria.

We recall that these fractions are computed with respect to the total
number of galaxies with $M_* \geq 7\times10^{10}M_\odot$ in our
sample. Since we are interested in the ratio between galaxy classes,
the impact of the exact choice of this threshold is measurable but not
dramatic. We verified that by either increasing or decreasing the mass limit
cut by a factor of 2, the fraction of both ``quiescent'' and
``red and dead'' galaxies in different redshift ranges changes by about
0.1, in the data (where applicable) as well as in the theoretical
models. This factor of 2 variation may also be produced by adopting a different IMF, and is
similar to the uncertainty in the stellar mass estimate.

Our analysis confirms the cosmological decrease in the number density of massive early--type
galaxies at high redshifts: quiescent galaxies dominate the
population of massive galaxies at $z<<1$, and become progressively
less common at higher $z$. However, we note that a
significant fraction of galaxies with low levels of SFR is in
place even at the highest redshifts sampled here ($z\simeq 3.5$) with
a fraction of about 10-15\% at $z>2$.

By definition, ``quiescent'', massive galaxies assembled most of
their stellar mass in previous epochs, implying that they experienced an active starburst phase or important merging
process at higher redshifts. The large fraction of ``quiescent''
galaxies that we observe at $z\simeq 2$ implies that these processes
must have been frequent at very high redshifts. The upper panel of
Fig. \ref{ActQuie} can also be interpreted in terms of ``active''
galaxies, i.e., the complementary galaxy fraction with $\dot{M}/M_*\geq
(t_U(z))^{-1}$. According to the duty-cycle argument presented in S09,
these actively star forming galaxies experienced a major episode
of star--formation, potentially building up a substantial fraction of
their stellar mass in this episode. Their fraction increases with
redshift, constituting more than 50\% of the massive galaxy population
at $z\geq 2$.

Finally, the sizeable number of ``red and dead'' galaxies is already in
place at $z>1.5$, implying that the star--formation episodes must be
quenched either by efficient feedback mechanism and/or by the
stochastic nature of the hierarchical merging process.

It is interesting to determine whether theoretical models agree with these
observational results.  In Fig. \ref{ActQuie}, we plot the predictions
of several models, applying the  same criteria based on
the $\dot{M}/M_*$ values.  We consider purely semi-analytical models
(\citet{menci06}, M06, \citet{fontanot07}, F07), a semi-analytical
rendition of the Millennium N-body dark matter simulation
(\citet{kitzbichler07}, K07), and purely hydrodynamical simulations
(\citet{nagamine06}, N06). The final are presented for three
different timescales $\tau$ of the star--formation rate (ranging from
$2\times10{^7}$ yrs to $2\times10{^8}$ yrs), and represented with a
shaded area. All these models agree in predicting a gradual decline
with redshift in the fraction of galaxies with a low SFR.

As far as the ``quiescent'' fraction is concerned, we note that most
models agree quantitatively in their predictions at all
redshifts. There is a general tendency to slightly overpredict the
fraction of ``quiescent'' galaxies, although the relatively large
error bars of our sample prevent firm conclusions from being drawn.  As we show in more
detail in S09 \citep[see also][]{daddi07}, this is the result of an
overall underestimate of the median SSFR of massive galaxies, which
increases the number of mildly star--forming galaxies that we
detect within our selection criteria $\dot{M}/M_*<(t_U(z))^{-1}$.  The
F07 models provides a noticeable exception to this process. A main
success of this model, indeed, is its capability of reproducing the
Scuba counts and the high associated SFR \citep{fontanot07}: it is unsurprising that
it also predicts a large fraction of active galaxies, and hence a
smaller fraction of ``quiescent'' ones.

Conversely, the predicted fraction of ``red and dead'' galaxies
varies significantly at all redshifts. This reveals that the
predicted fraction of galaxies with very low levels of SFR is a
particularly sensitive quantity, and provides a powerful
way of highlighting the differences between the models.  Some
models (M06, F07) underpredict the fraction of ``red and dead'' galaxies
at all redshifts, and in particular predict virtually no object at
$z>2$, in contrast to what observed. The Millennium-based model agrees
with the observed quantities, while the hydro model appears to
overpredict them.

It is beyond the scope of the present paper to discuss the origins of
these differences.  They are difficult to ascertain, because of
the complex interplay between all the physical processes involved in
these models, the different physical process implemented - most
notably those related to AGN feedback - and their different
technical implementations. The failure of most models to
reproduce simultaneously the fraction of ``quiescent'' and ``red and dead''
massive galaxies in the early Universe probably implies that the balance
between the amount of cool gas and the star--formation efficiency on
the one side, and the different feedback mechanisms on the other, is
still poorly understood.

\begin{acknowledgements}
  We are grateful to Mark Dickinson, Roberto Maiolino and Pierluigi
  Monaco for the useful discussions.  We also thank K. Nagamine for
  providing the output of his models. We are also in debt with the
  two referees for useful and prompt suggestions, that improved the
  presentation of the work. This work is based on observations carried
  out with the Very Large Telescope at the ESO Paranal Observatory
  under Program ID LP168.A-0485 and ID 170.A-0788 and the ESO Science
  Archive under Program IDs 64.O-0643, 66.A-0572, 68.A-0544,
  164.O-0561, 163.N-0210, and 60.A-9120. The Millennium Simulation
  databases used in this paper and the web application providing
  online access to them were constructed as part of the activities of
  the German Astrophysical Virtual Observatory. We acknowledge financial contribution from contract ASI I/016/07/0 (COFIS).
\end{acknowledgements}

\bibliographystyle{aa}
\bibliography{biblio}

\end{document}